\documentclass[3p,times]{elsarticle}
 
\usepackage{ecrc}
 
\volume{00}
 
\firstpage{1}
 
\journalname{Applied Surface Science}
 
\runauth{Chang et al.}
 
\jid{jnm}
 
\jnltitlelogo{Applied Surface Science}

\usepackage{lineno,hyperref} 
\modulolinenumbers[5] 
 
\usepackage{mathtools}
\usepackage[usenames, dvipsnames]{xcolor}
\usepackage{subcaption}

\usepackage{tikz}
\usepackage{tikz-qtree}
\usepackage{tkz-berge}
\usepackage{sansmath}
\usetikzlibrary{shadings,intersections}
\usetikzlibrary{shapes.geometric, arrows.meta}
\usepackage{tikz-3dplot}
\usetikzlibrary{calc,3d,decorations.markings,backgrounds,positioning,intersections,shapes,quotes,angles}
\usepackage{nicefrac}
\usepackage{caption}
\usepackage{subcaption}
\usepackage{mathptmx}
\usepackage{graphicx}
 
\usepackage[]{hyperref}
\hypersetup{
    colorlinks,%
    citecolor=blue,%
    filecolor=black,%
    linkcolor=blue,%
    urlcolor=black
}
 
\usepackage{cleveref}
\usepackage{tabularx}
\usepackage{multicol}
 
\biboptions{sort&compress}
 
\usepackage{scrextend}
 
\begin{document}
 
\begin{frontmatter} 
\title{Calculation of secondary electron emission yields from low-energy electron deposition in tungsten surfaces} 

\author[UCLA_MSE]{Hsing-Yin Chang}
\ead{irischang@ucla.edu}
\author[UCLA_MSE]{Andrew Alvarado}
\author[UCLA_MSE,UCLA_MAE]{and Jaime Marian}
 
\address[UCLA_MSE]{Department of Materials Science and Engineering, University of California, Los Angeles, CA 90095, USA}
\address[UCLA_MAE]{Department of Mechanical and Aerospace Engineering, University of California, Los Angeles, CA 90095, USA}
 
\begin{abstract}
We present calculations of secondary electron emission (SEE) yields in tungsten as a function of primary electron energies between 50 eV and 1 keV and incidence angles between 0 and 90$^{\circ}$. We conduct a review of the established Monte Carlo methods to simulate multiple electron scattering in solids and select the best suited to study SEE in high-Z metals. 
We generate secondary electron yield and emission energy functions of the incident energy and angle and fit them to bivariate fitting functions using symbolic regression.
We compare the numerical results with experimental data, with good agreement found. Our calculations are the first step towards studying SEE in nanoarchitected surfaces for electric propulsion chamber walls. \\
\end{abstract}
 

\end{frontmatter} 

 
\section{Introduction}
Secondary electron emission (SEE) is the emission of free electrons from a solid surface, which occurs when these surfaces are irradiated with external (also known as \emph{primary}) electrons. SEE is an important process in surface physics with applications in numerous fields, such as electric propulsion \cite{hobbs1967heat,raitses2005electron,raitses2006measurements,goebel2008fundamentals,mazouffre2016electric}, particle accelerators \cite{wiedemann2015particle}, plasma-walls in fusion reactors \cite{mccracken1979plasma,kaganovich2007kinetic,gunn2012evidence,lee2013secondary,post2013physics}, electron microscopy and spectroscopy \cite{seiler1983secondary,reimer2000scanning}, radio frequency devices \cite{kim1989rf,chabert2011physics,charles2012investigation}, etc. In Hall thrusters for electric propulsion, a key component is the channel wall lining protecting the magnetic circuits from the discharge plasma. These channel walls are a significant factor in Hall thruster performance and lifetime through its interactions with the discharge plasma. These interactions are governed by the sheath formed along the walls, and so the properties of the sheath determine the amount of electron energy absorbed by the wall, which in turn affects the electron dynamics within the bulk discharge \cite{hobbs1967heat,dunaevsky2003secondary,sydorenko2009breakdown,campanell2012instability}.  Furthermore, the energy imparted by the sheath to the ions within the discharge determines the impact energy and incident angle of ions upon the surface, thus affecting the amount of material sputtered and consequently the wall erosion rate \cite{sheehan2014effects,langendorf2015effect}. Thus, understanding how SEE affects sheath stability is crucial to make predictions of channel wall lifetime.

Recently, a new wall concept based nano-architected surfaces has been proposed to mitigate surface erosion and SEE \cite{wang2007suppression,pivi2008sharp,nistor2014multipactor,sattler2016engineered}. Demonstration designs based on high-Z refractory materials have been developed, including architectures based on metal nanowires and nanofoams \cite{baglin2000secondary,suetsugu2010experimental,aguilera2013cuo,raitses2013electron,patino2016secondary}. The idea behind these designs is to take advantage of very-high surface-to-volume ratios to reduce SEE and ion erosion by internal trapping and redeposition. Preliminary designs are based on W, W/Mo, and W/Re structures, known to have intrinsically low sputtering yields secondary electron emission propensity. 
A principal signature of electron discharges in plasma thrusters is the low primary electron energies expected in the outer sheath, on the order of 100 eV, and only occasionally in the several hundred eV regime. Accurate experimental measurements are exceedingly difficult in this energy range due to the limited thickness of the sheath layer, which is often outside the resolution of experimental probes \cite{staack2004shielded,sheehan2011comparison,sheehan2016recommended}. Modeling then suggests itself as a complementary tool to experiments to increase our qualitative and quantitative understanding of SEE processes.

To quantify the net SEE yield from these surfaces, models must account for the explicit geometry of these structures, which requires high spatial resolution and the capacity to handle large numbers of degrees of freedom. However a precursor step to the development of these descriptions is the characterization of the SEE yield functions as a function of incident electron energy and angle of incidence in flat surfaces. Once defined, these functions can then be implemented at the level of each surface element to create a spatially-dependent emission picture of the SEE process. This is the subject of the present paper: to calculate SEE yield functions from flat W surfaces in terms of primary electron energy and incidence angle. To this end, we carry out Monte Carlo calculations of electron scattering processes in pure W using a series of scattering models specifically tailored to high-Z metals.

The paper is organized as follows. First we discuss the theoretical models employed to study electron scattering in W. This is followed by a discussion of the implementation of these models under the umbrella of a Monte Carlo framework. Our results follow, with emphasis on emission yield and energy functions. We finalize with the conclusions and the acknowledgments.



\section{Theory and Methods} 
\subsection{Electron Scattering Theory}
The present model assumes that electrons travel in an isotropic continuum medium undergoing collisions with bulk electrons. Each collision results in a trajectory change with an associated energy loss, which depend on the nature of the electron-electron interaction. As well, collisions may result in secondary electron production.
We classify interactions into two broad categories: elastic and inelastic, each characterized by the corresponding collision mean free path and an angular scattering function. These processes are then simulated using a Monte Carlo approach, where collisions are treated stochastically and trajectories are tracked as a sequence of scattering events until the resulting secondary electrons are either thermalized  or emitted back from the surface.

Scattering theory provides formulas for the total and the differential scattering cross sections, from which the mean free path and polar scattering angle can be obtained, respectively.  %
Next, we provide a brief description of the essential theory behind each of the distinct collision  processes considered here. Our implementation accounts for the particularities of low-energy electron scattering in high-Z materials.


\subsection{Elastic Scattering}

Elastic scattering takes place between electrons and atomic nuclei, which --due to the large mass difference--  results in no net energy loss for the electron, only directional changes \cite{mott1929scattering}. 
A widely used electron-atom elastic scattering cross section is the screened Rutherford scattering cross section \cite{joy1995monte,reimer1995monte}, which provides a simple analytical form and is straightforward to implement into a Monte Carlo calculation. However, the screened Rutherford scattering is generally not suitable for low-energy electron irradiation of high-Z metals. 
In this work, we use an empirical total elastic scattering cross section proposed by Browning \textit{et al.} (1994), which is obtained via fitting to trends in tabulated Mott scattering cross section data set described by Czy{\.z}ewski \textit{et al.} \cite{czyzewski1990calculations} using the relativistic Hartree-Fock potential. This is amenable to fast Monte Carlo computations at a high degree of accuracy.
The equation for the total elastic scattering cross section is \cite{browning1994empirical,browning1995low}:
\begin{equation}
\sigma_{el} = \frac{3.0\times10^{-18}Z^{1.7}}{(E+0.005Z^{1.7}E^{0.5}+0.0007Z^{2}/E^{0.5})} \quad [\rm cm^{2}],
\end{equation}
which is valid for atomic numbers up to 92 and for energies from 100 eV to 30 keV.
From this, the elastic mean free path can be derived:
\begin{equation}
\lambda_{el} = \frac{1}{N \sigma_{el}} = \frac{A}{N_{a} \rho \sigma_{el}} \quad [\rm cm]
\end{equation}
where $N$ is the number of atoms per cm\textsuperscript{3}.
For its part, the polar scattering angle can be randomly obtained as:
\begin{equation}
R = \frac{\int_{0}^{\theta} \left(\frac{d\sigma_{R}}{d\Omega}\right) d\Omega}{\int_{0}^{\pi} \left(\frac{d\sigma_{R}}{d\Omega}\right) d\Omega}
\end{equation}
where $d\Omega=2\pi \sin\theta d\theta$ is the infinitesimal solid angle.

Solving the above equation for the Mott cross section requires numerical integration, as there is no simple analytical form for the polar scattering angle $\theta$. Drouin \cite{drouin1994computation} \textit{et al.} (1994) gives a parameterized form of the function as
\begin{equation}
\cos (\theta_{i}^{\beta}) = 1-\frac{2 \alpha R^{*}}{1+\alpha-R^{*}} \\
\end{equation}
where $\theta_{i}$ is given in degrees.
Then first parameter, $\alpha$, as a function of the energy is obtained with
\begin{equation}
\log_{10}(\alpha) = a+b\log_{10}(E)+c\log_{10}^{2}(E)+\frac{d}{e^{\log_{10}(E)}} 
\end{equation}
where $E$ is the energy in keV, $a, b, c$ and $d$ are constants that have been calculated using the least-square method, and $e = 2.7813$. A tabulation form of $a, b, c$ and $d$ for the first 94 elements of the periodic table is found in Table 2 in reference \cite{drouin1994computation}. 
For tungsten ($Z = 74$), $a = -2.0205, b = -1.2589, c = 0.271737, d = -0.695477$.

The second parameter, $\beta$, is calculated using the following equations:
\begin{equation}
\begin{aligned}
& \beta^{*} = a+b \sqrt{E} \ln(E)+\frac{c\ln(E)}{E}+\frac{d}{E} \\
&\beta = \left\{\begin{aligned}
& 1 \quad \text{if} \quad \beta^{*}>1 \\
& \beta^{*} \quad \text{if} \quad \beta^{*}\leq1
\end{aligned}\right.
\end{aligned}
\end{equation}
where $E$ is the energy in keV, $a, b, c$ and $d$ are constants that have been obtained using the least-squares fitting. A tabulation form of $a, b, c$ and $d$ for the first 94 elements of the periodic table is found in Table 3 in reference \cite{drouin1994computation}. 
For tungsten ($Z = 74$), $a = 0.71392, b = 0.00197916, c = -0.0172852, d = -0.0570799$.

Typically, the random number $R$ generated is uniformly distributed among 0 and 1 in single-scattering Monte Carlo program. This is why $R^{*}$, a modified version of $R$, uniformly distributed between 0 and the maximum value computed by $\cos (\theta_{i}^{\beta}) = 1-\frac{2 \alpha R^{*}}{1+\alpha-R^{*}}$ is used.
The maximum value $R_{max}$ is obtained when $\theta_{i}$ is set to 180$^{\circ}$. It is computed using this equation:
\begin{equation}
R_{max} = \frac{\cos(180^{\beta})+\alpha \cos(180^{\beta})-1-\alpha}{\cos(180^{\beta})-1-2 \alpha}
\end{equation}
$R^{*}$ is then obtained by the product of $R$ and $R_{max}$.
\begin{equation}
R^{*} = R \times R_{max}
\end{equation}

The azimuthal angle $\phi$ can take on any value in the range 0 to 2$\pi$ determined by a random number $R$ uniformly distributed in that range.
\begin{equation}
\phi = 2\pi R
\end{equation}

\subsection{Inelastic Scattering}
In contrast to elastic scattering, inelastic scattering implies collisional energy loss.  There are  several distinct inelastic interaction processes to be considered, including phonon excitation, secondary electron excitation, \emph{Bremsstrahlung} or continuum X-ray generation, and ionization of inner electron shells. Each mechanism is described by a model that provides expressions for the scattering cross section, scattering angle, and mean free path. The physics behind some of these processes is complex, and detailed expressions for the associated cross sections are often unavailable \cite{egerton2011electron,lin2007study}.

In conventional Monte Carlo approaches, Bethe's theory of stopping power based on a continuous slowing-down approximation (CSDA) \cite{bethe1930theorie,joy1995monte,dapor2014transport} is used to describe the average energy dissipation rate of a penetrating electron along its path, in which the contribution of all possible excitation processes to the energy loss has been represented by a factor called the mean ionization energy, $J$. However, this formula is not valid in the low energy regime (0.1-30 keV) or for high atomic number elements ($Z > $ 30). To resolve this, much effort has been devoted to modifying the Bethe formula, from which systematization of tabulated electron stopping powers for various elements and attempts to simplify the calculations have emerged. \cite{joy1989empirical,fernandez1993inelastic,jablonski2006new,taborda2015simple,jablonski2016predictive} In general, the use of these formulas for elements or compounds with fitting parameters requires a detailed and accurate supply of experimental data on which to base its physics and against which to test its predictions. \cite{joy1995database} Nevertheless, the CSDA strategy may still become obsolete when an electron occasionally loses a large fraction of its energy in a single collision as well as when secondary electron emission distribution spectra are required. To develop a more comprehensive Monte Carlo approach, incorporating differential cross sections for each of the inelastic events seems necessary \cite{shimizu1976monte,adesida1980study,shimizu1992monte,ding1996monte}. 

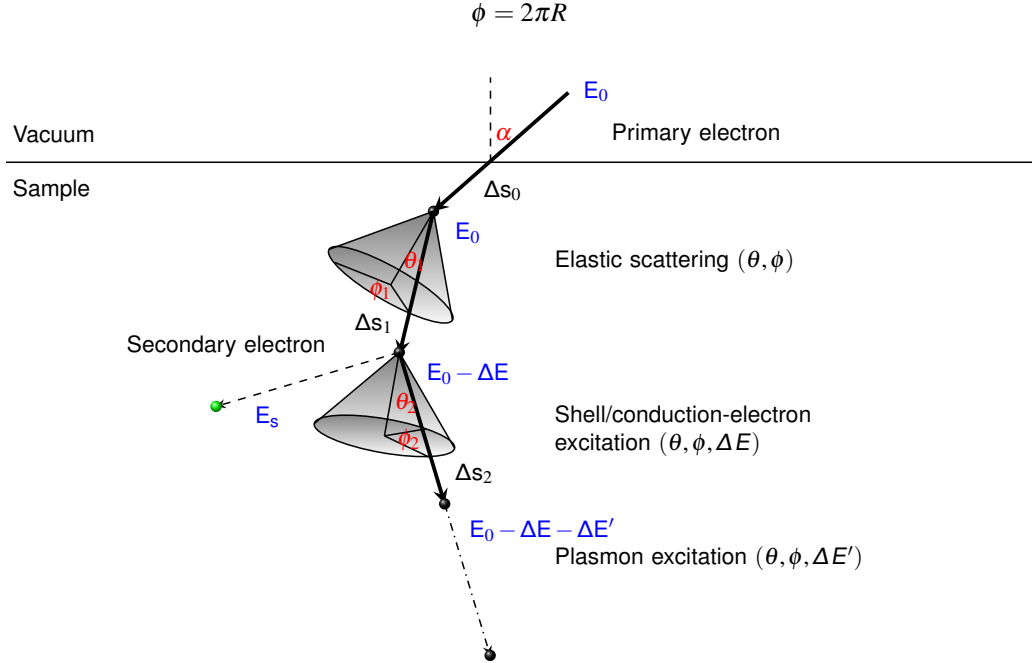
\begin{figure}[htb]
\begin{center}
\begin{tikzpicture}[thick,scale=0.75, every node/.style={font=\sffamily\sansmath\large, transform shape}]
\coordinate (O) at (0,0);

\draw[line width=0.2mm]  (-8,0) -- (10,0);
\draw[line width=0.2mm,dashed] (0.5,0) -- (0.5,1.5);
\begin{scope}[rotate=-30]
\shade[color = gray] (-1.25,-2.5) -- (0,-1) -- (1.25,-2.5);
\shade[color = gray] (0,-2.5) circle(1.25cm and 0.3cm);
\shade[ball color = black] (0,-1) circle (0.1cm);
\draw[line width=0.5mm,->,>=stealth] (1,2) -- (0,-1);
\draw[line width=0.5mm,->,>=stealth] (0,-1) -- (0.75,-3.5);
	\draw[line width=0.2mm] (0,-2.5) circle(1.25cm and 0.3cm);
	\draw[line width=0.2mm] (0,-2.5) -- (0,-1);
	\draw[line width=0.2mm] (0,-2.5) -- (0.5,-2.75);
	\draw[line width=0.2mm] (0,-2.5) -- (-1.05,-2.65);
	\draw[line width=0.2mm] (-1.25,-2.5) -- (0,-1) -- (1.25,-2.5);
\end{scope}
\begin{scope}[rotate=-10] 
\shade[color = gray] (-1.75,-5) -- (-0.5,-3.5) -- (0.75,-5);
\shade[color = gray] (-0.5,-5) circle(1.25cm and 0.3cm);
\draw[line width=0.5mm,->,>=stealth] (-0.5,-3.5) -- (0.75,-6);
\draw[line width=0.2mm,->,>=stealth,dashed] (-0.5,-3.5) -- (-3.5,-5);
\draw[line width=0.2mm,->,>=stealth,dash dot] (0.75,-6) -- (2,-8.5);
\shade[ball color = green] (-3.5,-5) circle (0.1cm);
\shade[ball color = black] (-0.5,-3.5) circle (0.1cm);
\shade[ball color = black] (0.75,-6) circle (0.1cm);
\shade[ball color = black] (2,-8.5) circle (0.1cm);
	\draw[line width=0.2mm] (-0.5,-5) circle(1.25cm and 0.3cm);
	\draw[line width=0.2mm] (-0.5,-5) -- (-0.5,-3.5);
	\draw[line width=0.2mm] (-0.5,-5) -- (0.4,-5.25);
	\draw[line width=0.2mm] (-0.5,-5) -- (0.2,-4.75);
	\draw[line width=0.2mm] (-1.75,-5) -- (-0.5,-3.5) -- (0.75,-5);
\end{scope}
\node[text width=8cm, anchor=east, left] at (0.25,0.5) {Vacuum};
\node[text width=8cm, anchor=east, left] at (0.25,-0.5) {Sample};
\node[text width=6cm, anchor=east, left] at (0.25,-3.25) {Secondary electron};
\node[text=red, text width=6cm, anchor=east, left] at (5.1,-1.75) {$\theta_{1}$};
\node[text=red, text width=6cm, anchor=east, left] at (4.5,-2.25) {$\phi_{1}$};
\node[text width=6cm, anchor=east, left] at (4.25,-2.85) {$\rm \Delta s_{1}$};
\node[text=red, text width=6cm, anchor=east, left] at (4.95,-4.25) {$\theta_{2}$};
\node[text=red, text width=6cm, anchor=east, left] at (5,-4.9) {$\phi_{2}$};
\node[text=blue, text width=6cm, anchor=east, left] at (2.5,-4.5) {$\rm E_{s}$};
\node[text=blue, text width=6cm, anchor=west, right] at (2,1.25) {$\rm E_{0}$};
\node[text=red, text width=6cm, anchor=west, right] at (0.45,0.5) {$\alpha$};
\node[text width=6cm, anchor=west, right] at (2.5,0.5) {Primary electron};
\node[text width=6cm, anchor=west, right] at (0.25,-0.5) {$\rm \Delta s_{0}$};
\node[text width=6cm, anchor=west, right] at (1.5,-1.75) {Elastic scattering $(\theta,\phi)$};
\node[text=blue, text width=6cm, anchor=west, right] at (-0.25,-1.25) {$\rm E_{0}$};
\node[text width=8cm, anchor=west, right] at (1.5,-4.75) {Shell/conduction-electron \\ excitation $(\theta,\phi,\Delta E)$};
\node[text=blue, text width=6cm, anchor=west, right] at (-0.75,-3.75) {$\rm E_{0}-\Delta E$};
\node[text width=6cm, anchor=west, right] at (-0.25,-5.5) {$\rm \Delta s_{2}$};
\node[text width=8cm, anchor=west, right] at (1.5,-7) {Plasmon excitation $(\theta,\phi,\Delta E')$};
\node[text=blue, text width=6cm, anchor=west, right] at (0,-6.5) {$\rm E_{0}-\Delta E-\Delta E'$};

\end{tikzpicture}
\end{center}
\caption{Schematic diagram of the discrete collision model of electron scattering simulated using Monte Carlo} \label{fig:figure1}
\end{figure}

Ritchie \textit{et al.} (1969) have demonstrated that the stopping power described by Bethe's formula is obtained by the summation of theoretical stopping powers for conduction electron, plasmon and L-shell electron excitations for aluminum. \cite{shimizu1992monte} Fitting (1974) \cite{fitting1974transmission} has also shown that this stopping power derived by Ritchie \textit{et al.} is in very good agreement with experimental investigation even in the energy range between 0.8 and 4 keV. Accordingly, the model of inelastic scatterings considered in the present approach are electron-conduction electron scattering, electron-plasmon scattering and electron-shell electron scattering as shown in Figure \ref{fig:figure1}.

\subsubsection{Inner Shell Electron Ionization}
The classical formalism of Gryzi{\'n}ski (1965) \cite{gryzinski1959classical,gryzinski1965classical,gryzinski1965two,gryzinski1965two1} has been adopted to describe inner-shell electron ionization. The differential cross section can be written as:
\begin{equation}
\begin{aligned}
\frac{d \sigma_{s}(\Delta E)}{d \Delta E} & = \frac{\pi e^{4}}{(\Delta E)^{3}} \frac{E_{B}}{E} \Big(\frac{E}{E+E_{B}} \Big)^{3/2} \Big(1-\frac{\Delta E}{E} \Big)^{E_{B}/(E_{B}+\Delta E)} \\
& \times \bigg\{ \frac{\Delta E}{E_{B}}\Big(1-\frac{E_{B}}{E}\Big)+\frac{4}{3}\ln\bigg[2.7+\Big(\frac{E-\Delta E}{E_{B}}\Big)^{1/2} \bigg]\bigg\}
\end{aligned}
\end{equation}
where $\Delta E$, $E$ and $E_{B}$ are the energy loss, the primary electron energy, and the mean electron binding energy, respectively. 

At each inelastic scattering event, the energy loss of the primary electron resulting from an inelastic scattering with the shell is determined using a uniform random number $R$ and by finding a value of $\Delta E$ which satisfies the relation
\begin{equation}
R = \int_{E_{B}}^{\Delta E}\frac{d\sigma_{s}(\Delta E')}{d\Delta E'} \frac{d\Delta E'}{\sigma_{s}}
\end{equation}
The integral is given by the approximate expression \cite{bentabet2006electrons}
\begin{equation}
\begin{aligned}
& \int_{\Delta E}^{E}\frac{d\sigma_{s}(\Delta E')}{d\Delta E'}d\Delta E' \\ 
& = \Big(\frac{\pi n_{s}e^{4}}{EE_{B}} \Big) \Big(\frac{E}{E+E_{B}} \Big)^{3/2} \Big(1-\frac{\Delta E}{E}\Big)^{1+(E_{B}/(E_{B}+\Delta E))} \\
& \times \bigg \{ \frac{\Delta E}{E_{B}}+\frac{2}{3}\Big(1-\frac{\Delta E}{E}\Big) \ln \bigg[ 2.7+\Big(\frac{E-\Delta E}{E_{B}}\Big)^{1/2} \bigg] \bigg \} \\
& \times \frac{E_{B}^{2}}{\Delta E^{2}} \quad (\Delta E \geq E_{B})
\end{aligned}
\end{equation}
where $n_{s}$ is the occupation number of electrons in the shell.

The total cross section of the inner electron excitation is obtained by integrating over all possible values of $\Delta E$
\begin{equation}
\begin{aligned}
\sigma_{s}(E) & = \int_{E_{B}}^{\Delta E_{max}}\frac{d\sigma_{s}(\Delta E')}{d\Delta E'} d\Delta E' \\
& = 6.5141\times10^{-14}\frac{n_{s}}{E_{B}^2} \frac{E_{B}}{E} \Big(\frac{E-E_{B}}{E+E_{B}} \Big)^{3/2} \\
& \times \bigg[ 1+\frac{2}{3}\Big(1-\frac{E_{B}}{2E}\Big) \ln \Big(2.7+\Big(\frac{E}{E_{B}}-1 \Big)^{1/2}\Big) \bigg] [\rm cm^{2}]
\end{aligned}
\end{equation}
where the maximum amount of energy that can be lost $\Delta E_{max}$ is equal to E.

When the random number selection gives an energy loss less than the binding energy $E_{B}$, the actual energy loss is set to be zero. The scattering angle for an inelastic electron-electron event is calculated according to the binary collision approximation (BCA) as
\begin{equation}
\sin \theta = \Big(\frac{\Delta E}{E}\Big)^{1/2}
\end{equation}
In tungsten, for primary energies $E \leq$ 1keV, inner shell electron ionization can be safely neglected, as the energy is insufficient to knock out inner shell electrons.

\subsubsection{Conduction Electron Excitation}
For metals bombarded by electrons, Streitwolf (1959) \cite{stolz1959theorie} has given the differential cross section for conduction electron excitation by using perturbation theory as
\begin{equation}
\frac{d\sigma_{c}(E_{SE})}{dE_{SE}} = \frac{e^{4}N_{a}\pi}{E(E_{SE}-E_{F})^{2}}
\end{equation}
The total energy loss cross section $\sigma_{c}(E)$ can be obtained by integrating the above expression between the lower energy limit $E_{F}+\Phi$ and the upper energy limit $E$:
\begin{equation}
\sigma_{c}(E_{SE}) = \frac{e^{4}N_{a}\pi}{E} \frac{E-E_{F}-\Phi}{\Phi(E-E_{F})}
\end{equation}
The obtained relation samples the energy of the secondary electron with the random number $R$:
\begin{equation}
E_{SE}(R) = \Big[R E_{F}-A(E_{F}+\Phi) \Big]/(R-A)
\end{equation}
where $\Phi$ is the workfunction and $A$ = ($E$ $-$ $E_{F}$)/($E$ $-$ $E_{F}$ $-$ $\Phi$).
Once the energy of the secondary electron is known (equal to the energy lost by the primary electron), the next question is how the two electrons are oriented in space. More accurate results can be obtained if the classical BCA is used, which results from conservation of energy and momentum. The azimuthal angle is again assumed to be isotropic. For the incident electron, we then have:
\begin{equation}
\sin \theta = \sqrt{\frac{\Delta E}{E}}
\end{equation}
\begin{equation*}
\phi = 2\pi R
\end{equation*}
where $\Delta E$ is the energy lost by the incident electron.
For the secondary electrons, scattering angles can be calculated as follow:
\begin{equation}
\sin \vartheta = \cos \theta
\end{equation}
\begin{equation}
\varphi = \pi+\phi
\end{equation}
The above expression is applied to the inner shell electron as well as conduction electron excitations.

\subsubsection{Plasmon Excitation}
The Coulomb field of the primary electron can perturb electrons of the solid at relatively long range as it passes through the target. The primary electron can excite oscillations (known as \emph{plasmons}) in the conduction electron gas that exists in a metallic sample with loosely bound outer shell electrons. The differential cross section for plasmon excitation is given by Ferrel (1956) \cite{ferrell1956angular,ferrell1957characteristic,joy1986principles}, per conduction-band electron per unit volume
\begin{equation}
\frac{d \sigma_{p}(E,\theta)}{d \Omega} = \frac{1}{2 \pi a_{0}} \frac{\theta_{p}}{\theta^2+\theta_{p}^2}
\end{equation}
\begin{equation}
\theta_{p} = \frac{\Delta E}{2E} = \frac{\hslash \omega_{p}}{2E}
\end{equation}
where 
$a_{0}$ is Bohr radius ($5.29\times10^{-9}$ [cm]).
In plasmon scattering, primary electron energy loss is quantized and ranges from 3 to 30 eV depending on the target species, which is detected as strong features in electron energy-loss spectra (EELS). Plasmon scattering is so sharply peaked forward that the total plasmon cross section, $\sigma_{p}$, can be found by setting $d\Omega = 2\pi \sin \theta d\theta \approx 2\pi \theta d\theta$:
\begin{equation}
\sigma_{p} = \int d\sigma_{p}(\theta) = \frac{\theta_{p}}{2\pi a_{0}}\int_{0}^{\theta_{1}} \frac{2\pi \theta d\theta}{\theta^2+\theta_{p}^2}
\end{equation}
By assuming the upper integration limit as $\theta_{1}$ = 0.175 rad, where $\theta \approx \sin\theta$, and incorporating the factor $(n_{c}A/N_{a}\rho)$ to put the cross section on a per-atom/$\rm cm^{2}$ basis gives the total cross section of the plasmon excitation as
\begin{equation}
\sigma_{p} = \frac{n_{c}A\theta_{p}}{2N_{a}\rho a_{0}} \Big[\ln(\theta_{p}^{2}+0.175^{2})-\ln(\theta_{p}^{2}) \Big] \quad [\rm cm^{2}]
\end{equation}
where $n_{c}$ is the number of conduction-band electrons per atom.
Essentially, the scattering of primary electrons due to plasmon excitations is restricted with $\theta < \theta_{max}$, $k_{c}$ being the cut-off wavenumber. Since $\theta_{max}$ is so small, about 10 mrad in the energy range discussed here, the angular deflection due to plasmon excitation is neglected in this approach.

Again, the azimuthal angle $\phi$ can take on any value in the range 0 to 2$\pi$ selected by a random number $R$ uniformly distributed in that range.
\begin{equation*}
\phi = 2\pi R
\end{equation*}

\section{Monte Carlo Calculations}
As indicated above, electron trajectories are simulated by generating a spatial sequence of collisions by randomly sampling from among all possible scattering events. The distance traveled by electrons in between collisions, $\Delta s$, is assumed to follow a Poisson distribution defined by the total mean free path $\lambda_T$ \cite{joy1995monte}
\begin{equation}
\Delta s = - \lambda_{T} \log R
\end{equation}
where
\begin{equation}
\frac{1}{\lambda_{T}} = \frac{1}{\lambda_{el}}+\frac{1}{\lambda_{p}}+\frac{1}{\lambda_{c}}+\frac{1}{\lambda_{s}}= N(\sigma_{el}+\sigma_{p}+\sigma_{c}+\sigma_{s}) ,
\end{equation} 
$N$ is the number of atoms per cm\textsuperscript{3} and $R$ is a random number uniformly distributed in the interval $(0,1]$.
From this, we define the following probabilities:
\begin{equation}
\begin{aligned}
& P_{el} = \lambda_{T}/\lambda_{el}: \text{the probability that the next collision will be elastic} \\
& P_{p} = \lambda_{T}/\lambda_{p}: \text{the probability that the next collision will cause a plasmon excitation} \\
& P_{c} = \lambda_{T}/\lambda_{c}: \text{the probability that the next collision will cause a conduction electron excitation} \\
& P_{s} = \lambda_{T}/\lambda_{s}: \text{the probability that the next collision will cause an inner shell electron excitation}
\end{aligned}
\end{equation}
The type of collision is then chosen based on the following partition of the value of $R$:.
\begin{equation}
\begin{aligned}
& 0 < R \leq P_{el} \Longrightarrow \text{elastic scattering} \\
& P_{el} < R \leq P_{el}+P_{p} \Longrightarrow \text{plasmon excitation} \\
& P_{el}+P_{p} < R \leq P_{el}+P_{p}+P_{c} \Longrightarrow \text{conduction electron excitation} \\
& P_{el}+P_{p}+P_{c} < R \leq 1 \Big( \equiv P_{el}+P_{p}+P_{c}+P_{s} \Big) \Longrightarrow \text{inner shell electron excitation}
\end{aligned}
\end{equation}
The flow diagram corresponding to the implementation of the model just described is provided in \ref{app:flow}.
Following this approach, electron trajectories\footnote{Trajectories are generated by stitching together each sequence of discrete collision events and referring each collision point to a global \emph{laboratory} frame of reference. The coordinate transformation employed here can be found in ref.~\cite{heinrich1976use} and is provided in \ref{app:coord}.} are tracked in the energy-position space until a scattered electron either thermalizes, i.e.~its energy follows below the surface escape threshold (Fermi level plus workfunction) within the material, or reaches the surface with a velocity having a component pointing along the surface normal with an energy larger than the escape threshold. In the latter case, the electron is tallied as a secondary electron and its energy and exit angle are recorded.

Next we analyze the Monte Carlo calculations performed following this method and present results of secondary electron yield and emission energies as a function of  primary electron energy and angle of incidence.

\section{Results}
\begin{figure}[h]
\begin{subfigure}{.49\textwidth}
  \centering
  \includegraphics[width=1\linewidth]{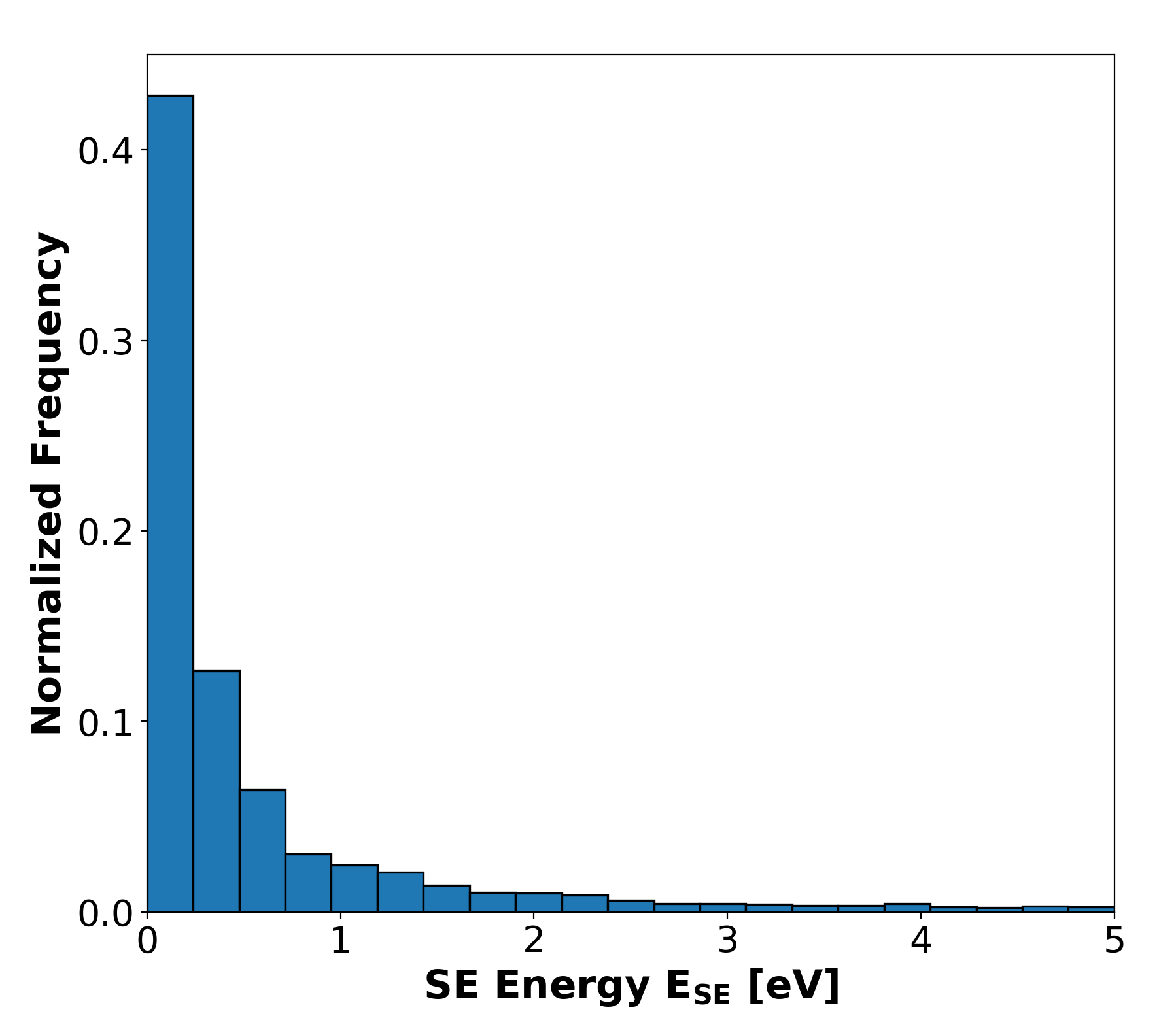}
  \caption{}
  \label{fig:sfig1}
\end{subfigure}
\begin{subfigure}{.49\textwidth}
  \centering
  \includegraphics[width=1\linewidth]{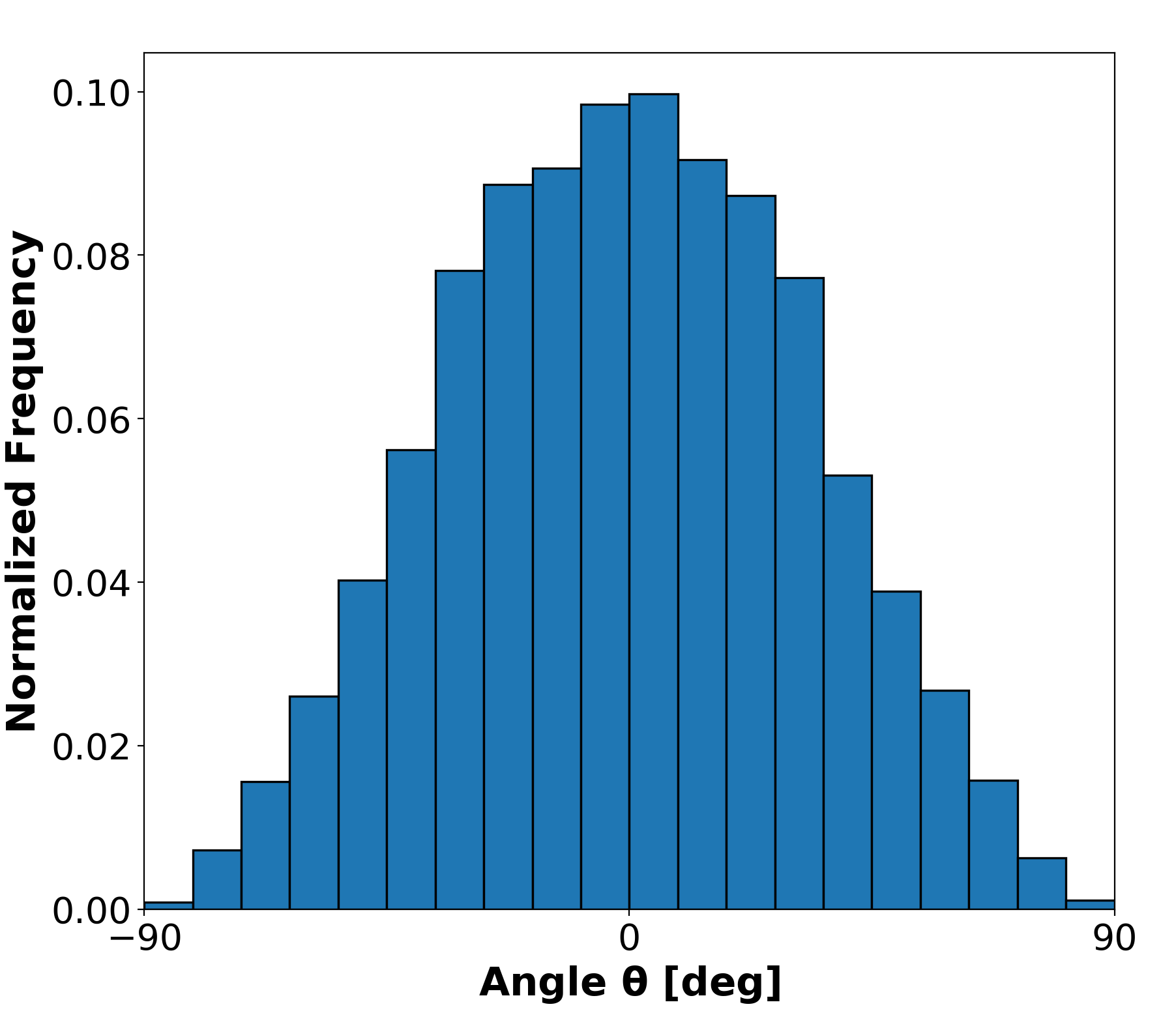}
  \caption{}
  \label{fig:sfig2}
\end{subfigure}
\caption{(a) Normalized energy distribution of secondary electron for 100 eV primary electron incident at 0$^{\circ}$. (b) Normalized angular distribution of secondary electron for 100 eV primary electron incident at 0$^{\circ}$.}
\label{fig:figure2}
\end{figure}

\begin{figure} 
\begin{center}
\includegraphics[scale=0.4]{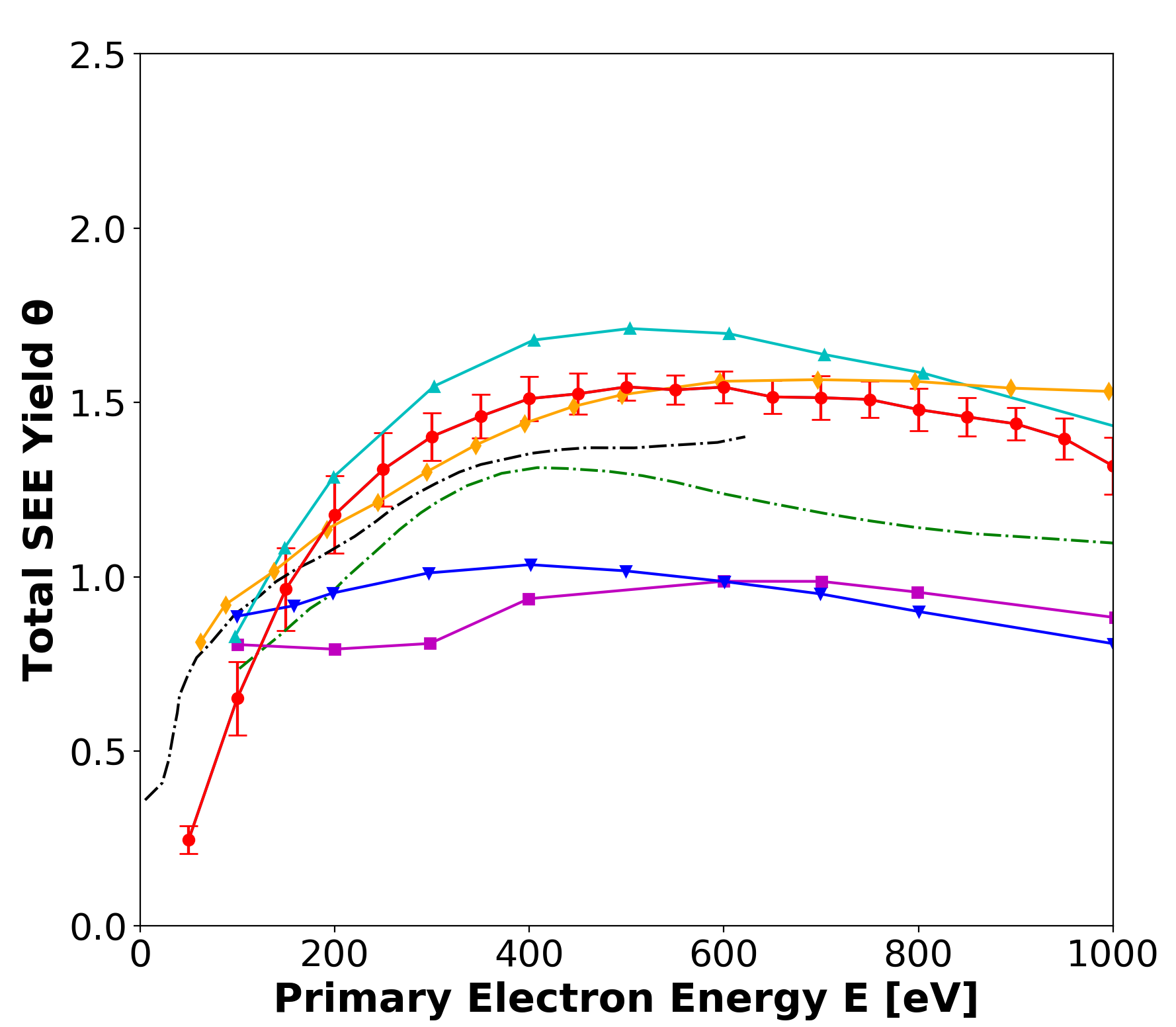}
\caption{The total SEE yield from smooth W as a function of primary electron energy, for electrons incident at 0$^{\circ}$. $\bullet$ = this work; black dash-dotted line = Ahearn (1931) \cite{ahearn1931emission}; green dash-dotted line = Coomes (1939) \cite{coomes1939total}; $\blacksquare$ = Bronshtein and Fraiman (1969) \cite{bronshtein1969honchnaya}; $\blacktriangle$ = Ding \textit{et al.} (2001) using the method where the SE is assumed to come from a distribution of electron energies in the valence band; $\blacktriangledown$ = Ding \textit{et al.} (2001) using the method where the SEs are assumed to originate from the Fermi level \cite{ding2001monte,walker2008secondary}; $\blacklozenge$ = Patino \textit{et al.} (2016) \cite{patino2016secondary}.}
\label{fig:figure3}
\end{center}
\end{figure}

\begin{figure} 
\begin{center}
\includegraphics[scale=0.4]{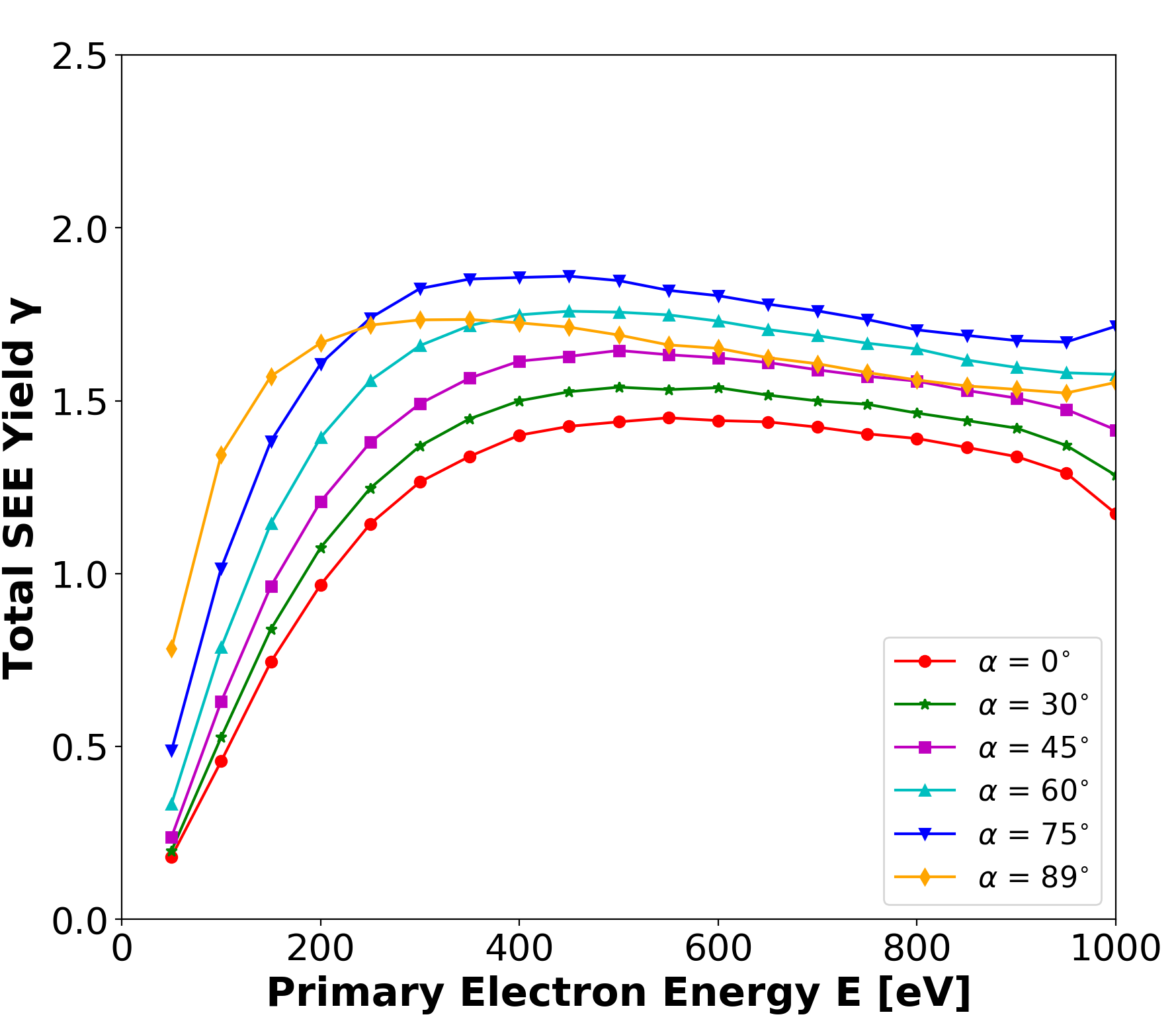}
\caption{The total SEE yield from smooth W as a function of primary electron energy, for electrons incident at 0$^{\circ}$, 30$^{\circ}$, 45$^{\circ}$, 60$^{\circ}$, 75$^{\circ}$ and 89$^{\circ}$.}
\label{fig:figure4}
\end{center}
\end{figure}

The total secondary electron yield for tungsten is calculated for primary beams at incident angles of 0$^{\circ}$, 30$^{\circ}$, 45$^{\circ}$, 60$^{\circ}$, 75$^{\circ}$ and 89$^{\circ}$ measured off the surface normal, for incident energies in the range 50-1000 eV. Two examples of the output from our code are shown in Figure \ref{fig:figure2}. Figure \ref{fig:sfig1} shows the normalized SEE energy distribution for an incident energy of 100 eV and normal incidence, while Figure \ref{fig:sfig2} gives the secondary electron angular distribution. The results are given in Figure ~\ref{fig:figure2}, \ref{fig:figure3} and \ref{fig:figure4}. The simulation results are found to agree reasonably well with experimental data.

\begin{figure}[h]
\begin{subfigure}{.49\textwidth}
  \centering
  \includegraphics[width=1\linewidth]{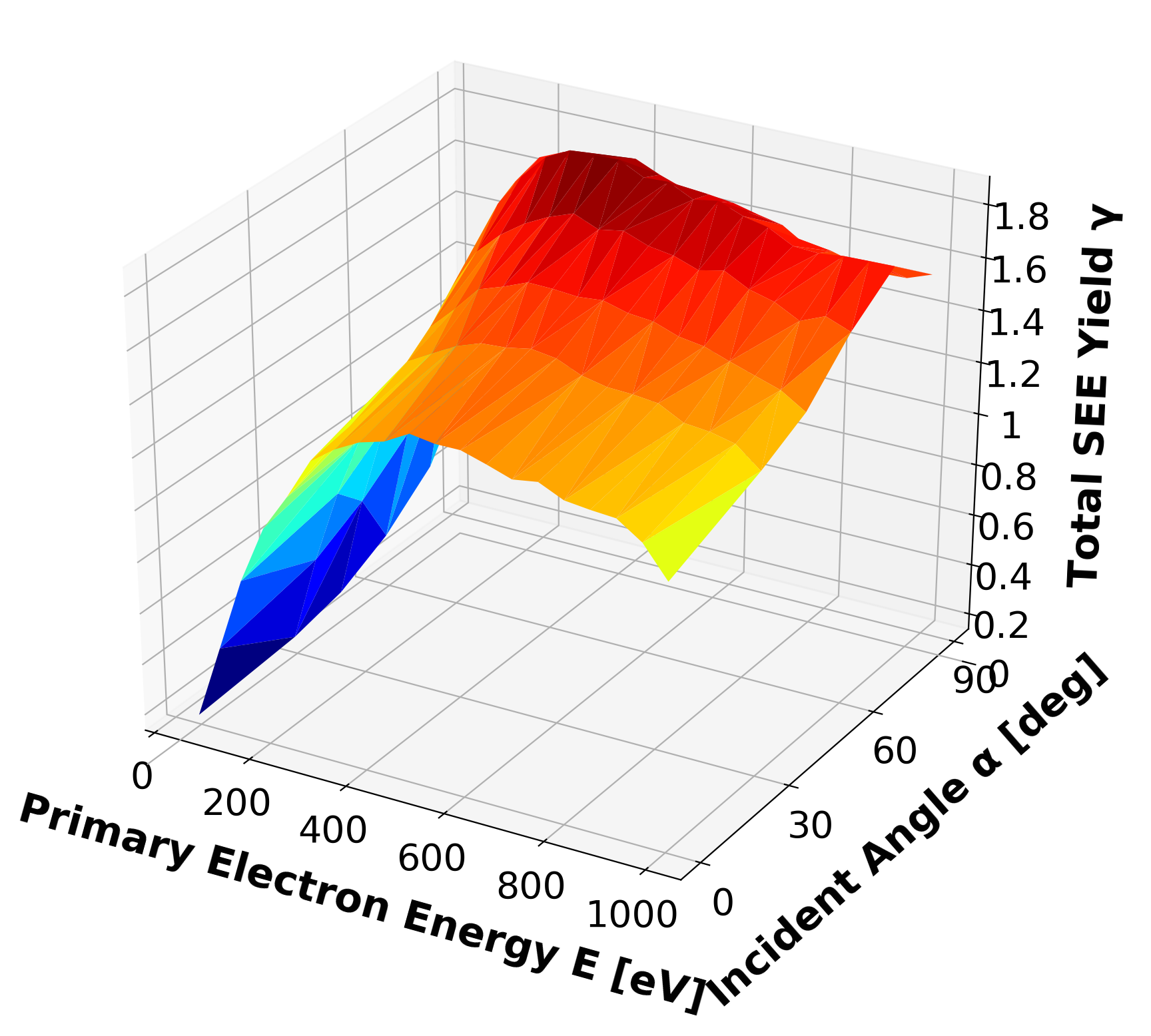}
  \caption{}
  \label{fig:sfig1}
\end{subfigure}
\begin{subfigure}{.49\textwidth}
  \centering
  \includegraphics[width=1\linewidth]{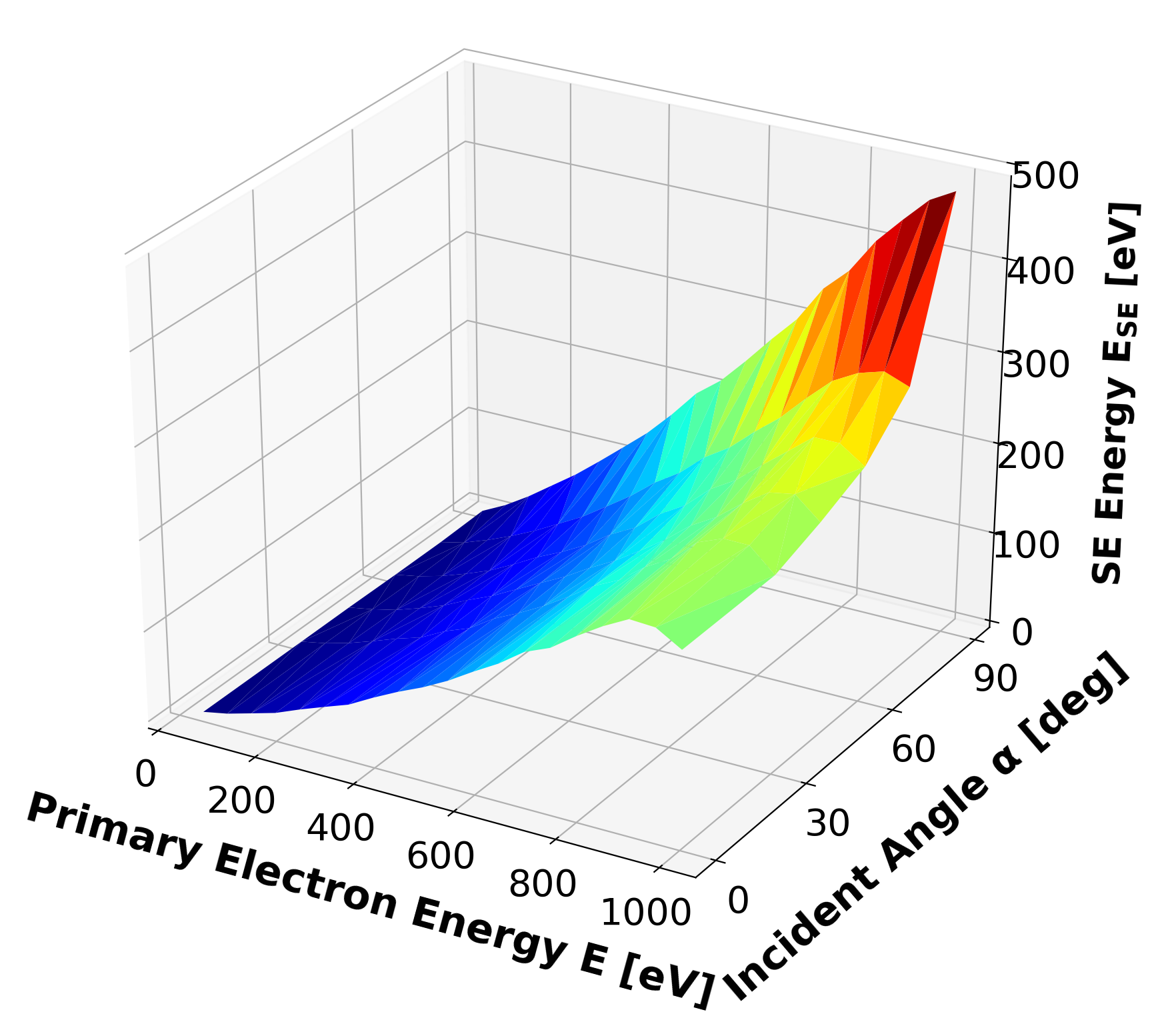}
  \caption{}
  \label{fig:sfig2}
\end{subfigure}
\caption{(a) Surface plot of the total SEE yield from smooth W as a function of primary electron energy and angle of incidence. (b) Surface plot of the secondary electron energy spectrum from smooth W as a function of primary electron energy and angle of incidence.}
\label{fig:figure5}
\end{figure}

The current results will be used in ray-tracing Monte Carlo simulations of SEE in arbitrary surface geometries. In these simulations, primary rays are generated above the material surface with the corresponding incident energy $E$. Intersections of these primary rays with surface elements determines the corresponding angle of incidence $\alpha$. $E$ and $\alpha$ are then used to sample from the data shown in, e.g., Figure ~\ref{fig:figure2}, after which secondary rays with appropriate energies $E_{SE}$ and exit angles (sampled from a cosine distribution) are generated. These daughter rays are themselves tracked in their interactions with other surface elements, after which the sequence is repeated and granddaughter rays are produced. This process goes on until rays either escape the surface with an upward velocity --in which case the event is tallied as a successful SEE event-- or until their energy is below the threshold escape energy (Fermi level plus workfunction). However, directly interpolating from our data tables potentially hundreds of thousands of times can slow down the simulations considerably. To avoid that, it is more efficient to fit the data to suitable analytical expressions that can be evaluated very fast on demand.
To this end, we fit our raw data to bivariate mathematical functions obtained using \emph{symbolic regression} (SR) via genetic evolutionary algorithms for machine learning \cite{schmidt2009distilling,schmidt2013eureqa}. SR utilizes evolutionary searches to determine both the parameters and the form of mathematical equations simultaneously.
The final expressions for the total SEE yield and energy distributions are 
\begin{equation}
\begin{aligned}
\gamma(E,\alpha) & = 1.91\times 10^{-2}E+1.3\times 10^{-4}\alpha^{3}+1.35\times 10^{-5}E^{2}-0.637 \\
& -4.46\times 10^{-12}\alpha^{5}E^{1/2}-9.68\times 10^{-4}E(E+0.114\alpha)^{1/2}
\end{aligned}
\end{equation}
\begin{equation}
\begin{aligned}
E_{\rm SE}(E,\alpha) & = 7.041313\times 10^{-2}E+1.7672\times 10^{-4}E^{2} \\
& +(0.1356545E+6.269578\times 10^{-5}E^{2} \\
& +\sin(1.767287\times 10^{-4}E^{2}))/(5.337308-5.241232\times 10^{-2}\alpha) \\
& -\sin(7.041313\times 10^{-2}E+1.767287\times 10^{-4}E^{2}) \\
& -0.13565454\alpha-4.050685\times 10^{-6}\exp(1.6751\times 10^{-2}E).
\end{aligned}
\end{equation}
We note that these expressions do not necessarily reflect the physics behind SEE and are just intended for efficient numerical evaluations strictly in the ranges shown in the figures. 


\section{Discussion and Conclusions}
Electron-matter interactions are complex processes. To make the theoretical treatment of electron scattering a tractable analytical problem, it is assumed that elastic scattering occurs through the interatomic potential, while inelastic scattering only through electron-electron interactions. Evidently, the accuracy of the Monte Carlo simulations depends directly on how precisely the approximations introduced in the model are described. Most models treat elastic interactions within Mott's formalism \cite{mott1929scattering} (or adaptations thereof). For their part, inelastic scattering processes in this work are considered individually, each one characterized by its own differential cross sections, corresponding to valence, inner shell, conduction, and plasmon electron excitation. 
In contrast, Ding \textit{et al.} \cite{ding2001monte} use Penn's dielectric function \cite{penn1987electron} for electron inelastic scattering obtained from a modification of the statistical approximation.
Many other models for metals account for valence interactions only \cite{koshikawa1974monte,kotera1989monte,joy1995monte}.

What is not captured in this work is generation of SE from plasmon decay, backscattered electrons, reflected electrons, and transmitted electrons (coming out from the back side of the sample). 
Comparison of the simulation results with experimental data provides a first-order check about the importance of these processes, which helps us understand the governing physics behind SEE. This is the purpose behind the comparison in Figure \ref{fig:figure3}.

%
%

To summarize, in this work we have carried out Monte Carlo calculations of low energy electron induced SE emission from flat tungsten surfaces. Our model includes multiple elastic and inelastic scattering processes, implemented via a discrete energy loss approach. We compare predictions of our model with other Monte Carlo techniques as well as experimental data, with generally good agreement found.
We have calculated the total SEE yield and secondary electron energy spectrum for primary electron beams at incident angles of 0$^{\circ}$, 30$^{\circ}$, 45$^{\circ}$, 60$^{\circ}$, 75$^{\circ}$ and 89$^{\circ}$, in the range 50-1000 eV. 
We have used SR to obtain analytical expressions that represent the numerical data. These functions are currently being used in ray-tracing Monte Carlo simulations of SEE in arbitrary  surface geometries.

\section*{Acknowledgement}
The authors acknowledge support from the Air Force Office of Scientific Research (AFOSR), through award number FA9550-11-1-0282 with UCLA.	

\bibliography{bibfile}

\appendix
\renewcommand\thefigure{\arabic{figure}}   
 
\label{key}
\section{List of Symbols}
\begin{tabular}{ l  l }
  $a_{0}$ & Bohr radius \\
  $A$ & atomic weight \\
  $Z$ & atomic number \\
  $N_{a}$ & Avogadro's number \\
  $\rho$ & density of the target \\
  $N=\rho N_{a}/A$ & atomic number density \\
  $n_{s}$ & number of electrons in shell or subshell \\
  $n_{c}$ & number of conduction-band electrons per atom \\
  $e$ & electron charge \\
  $\epsilon$ & permittivity of vacuum \\
  $m_{e}$ & mass of electron \\
  $\hslash$ & reduced Planck constant \\
  $\omega_{p}$ & plasma frequency \\
  $E_{F}$ & Fermi energy \\
  $k_{F}$ & Fermi wave number \\
  $E_{B}$ & binding energy of the shell \\
  $E_{pl}=\hbar \omega_{p}$ & plasmon energy \\
  $\Delta E$ & energy loss of primary electron \\
  $\Phi$ & work function \\
  $J$ & mean ionization potential \\
  $\alpha$ & incident angle of primary electron \\
  $\sigma_{el}$ & elastic scattering cross section \\
  $\sigma_{p}$ & plasmon excitation cross section \\
  $\sigma_{c}$ & conduction electron ionization cross section \\
  $\sigma_{s}$ & inner shell electron ionization cross section of \\
  d$\sigma$/d$\Omega$ & differential scattering cross section with respect to direction \\
  d$\sigma$/dE & differential scattering cross section with respect to energy \\
  $\lambda_{el}$ & elastic mean free path \\
  $\lambda_{p}$ & plasmon excitation mean free path \\
  $\lambda_{c}$ & conduction electron excitation mean free path \\
  $\lambda_{s}$ & inner shell electron excitation mean free path \\
  $\lambda_{T}$ & total mean free path \\
  $\theta$ & polar scattering angle of the primary electron \\
  $\vartheta$ & polar scattering angle of the secondary electron \\
  $\phi$ & azimuthal scattering angle of the primary electron \\
  $\varphi$ & azimuthal scattering angle of the secondary electron \\
  $\theta_{p}$ & plasmon loss scattering angle \\
\end{tabular}

\section{Constants \& Kinematical Quantities}
\begin{equation*}
\begin{aligned}
&N_{a} = 6.022\times10^{23}, \text{Avogadro's number} \\
&\epsilon = 8.85\times10^{-12} [\rm F/m], \text{permittivity of vacuum} \\
&m_{e} = 9.1\times10^{-31} [\rm kg] \\
&e = 1.6\times10^{-19} [\rm C], electron charge \\
&\hbar = 6.58\times10^{-16} [\rm eV\cdot s/rad], \text{reduced Planck constant} \\
&E_{h} = m_{e}e^{4}/\hbar^{2} = 2Ry = 27.2114 [\rm eV], \text{Hartree energy} \\
&Ry = 13.6 [\rm eV], \text{Rydberg energy} \\
&a_{0} =\hbar^{2}/(m_{e}e^{2}) = 5.29177\times10^{-9} [\rm cm], \text{Bohr radius} \\
&\pi e^{4} = \pi(a_{0}E_{h})^{2} = 6.5141\times10^{-14} [\rm cm^{2} \rm eV^{2}] \\
&m_{e} c^{2} = 510.999 [\rm keV], \text{rest energy of the electron} \\
\end{aligned}
\end{equation*}
The Fermi energy can be estimated using the number of electrons per unit volume as
\begin{equation*}
E_{F} = 3.64645\times10^{-15} n^{2/3} [\rm eV] = 1.69253 {n_{0}}^{2/3} [\rm eV]
\end{equation*}
where $n$ and $n_{0}$ are in the units of [cm$^{-3}$] and $n = n_{0} \times 10^{22}$.
The Fermi wave number is calculated as
\begin{equation*}
k_{F} = 6.66511\times10^{7} n_{0}^{1/3} [\rm cm^{-1}].
\end{equation*}
The Fermi velocity is calculated as
\begin{equation*}
v_{F} = 7.71603\times10^{7} n_{0}^{1/3} [\rm cm/s].
\end{equation*}

\newpage
\section{Program Flowchart} \label{app:flow}
\tikzset{%
  >={Latex[width=2mm,length=2mm]},
            base/.style = {rectangle, rounded corners, draw=black,
                           minimum width=4cm, minimum height=1cm,
                           text centered, font=\sffamily},
             switch/.style = {diamond, draw=black,
                           minimum width=4cm, minimum height=1cm,
                           text centered, font=\sffamily},                           
  activityStarts/.style = {base, fill=none},
       startstop/.style = {base, fill=none},
    activityRuns/.style = {base, fill=none},
         process/.style = {base, minimum width=2.5cm, fill=none,
                           font=\ttfamily},
         decision/.style = {switch, minimum width=2.5cm, inner sep=0pt, fill=none,
                           font=\ttfamily},
}
\tikzstyle{line} = [draw, -latex']

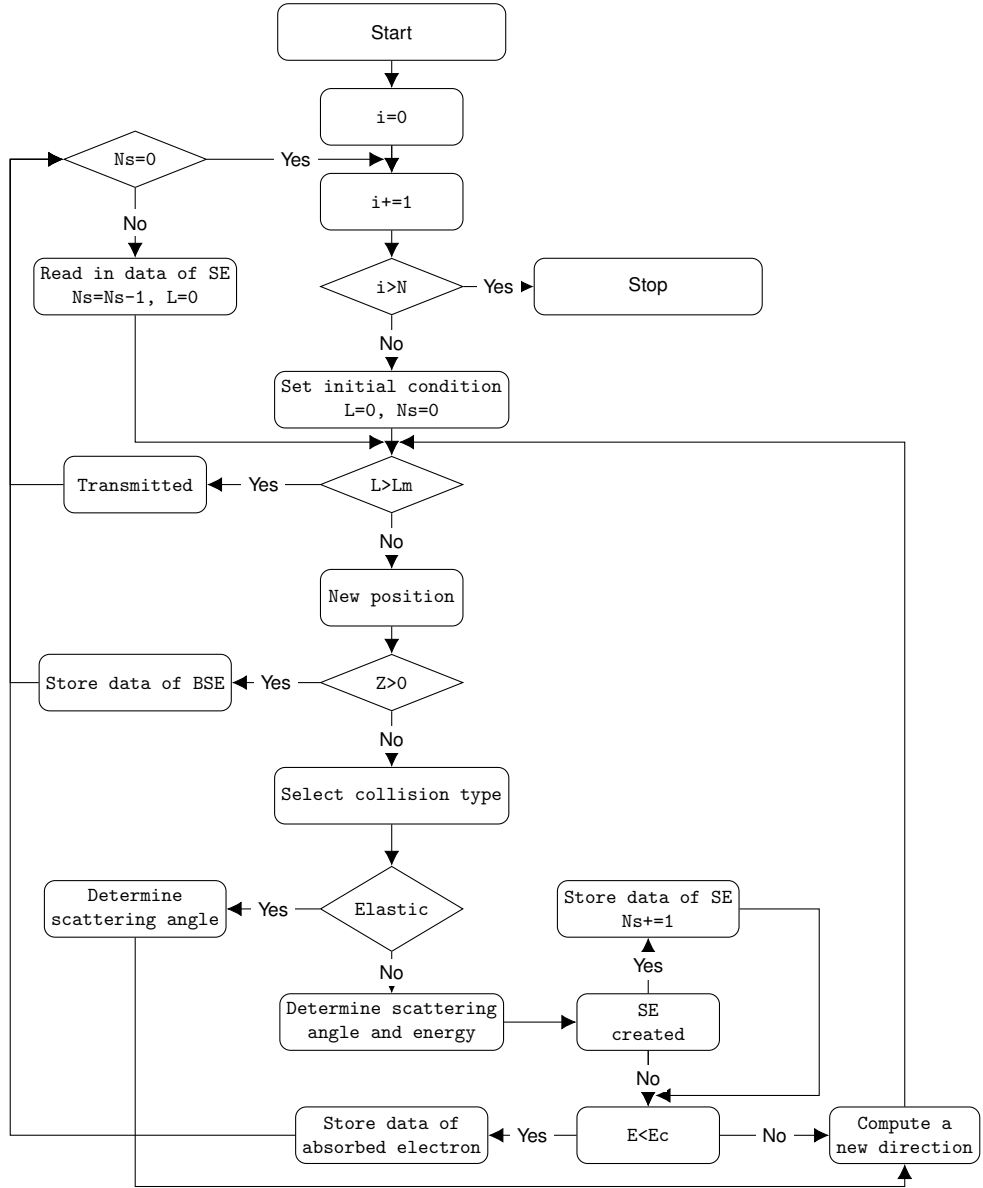
\begin{figure}[h]
\centering
\begin{tikzpicture}[scale=0.75, node distance=1.5cm,
    every node/.style={fill=white, font=\sffamily, transform shape}, align=center]
  \node (start)    [activityStarts]    {Start};
  \node (initial)    [process, below of=start]    {i=0};
  \node (next_0)    [process, below of=initial]    {i+=1};
  \node (return)    [decision, left of=initial, xshift=-3cm, yshift=-0.75cm]    {Ns=0};
  \node (decision_0)    [decision, below of=next_0]    {i>N};
  \node (stop)    [startstop, right of=decision_0, xshift=3cm]    {Stop};   
  \node (return2zero)    [process, left of=decision_0, xshift=-3cm]    {Read in data of SE\\ Ns=Ns-1, L=0};                                                     
  \node (set)    [process, below of=decision_0, yshift=-0.5cm]    {Set initial condition\\ L=0, Ns=0};
  \node (decision_1)    [decision, below of=set]    {L>Lm};
  \node (transmitted)    [process, left of=decision_1, xshift=-3cm]    {Transmitted};                                                     
  \node (position_new)    [process, below of=decision_1, yshift=-0.5cm]    {New position}; 
  \node (decision_2)    [decision, below of=position_new]    {Z>0};                                                   
  \node (backscattered)    [process, left of=decision_2, xshift=-3cm]    {Store data of BSE};                                                         
  \node (type)    [process, below of=decision_2, yshift=-0.5cm]    {Select collision type};     
  \node (elastic)    [decision, below of=type, , yshift=-0.5cm]    {Elastic};
  \node (angle)    [process, left of=elastic, xshift=-3cm]    {Determine\\ scattering angle};                                                              
  \node (SEEyield)    [process, right of=elastic, xshift=3cm]    {Store data of SE\\ Ns+=1};                                                    
  \node (angleANDenergy)    [process, below of=elastic, yshift=-0.5cm]    {Determine scattering\\ angle and energy};
  \node (SE)    [process, right of=angleANDenergy, xshift=3cm]    {SE\\ created};                                                              
  \node (absorbed)    [process, below of=angleANDenergy, yshift=-0.5cm]    {Store data of\\ absorbed electron};   
  \node (threshold)    [process, right of=absorbed, xshift=3cm]    {E<Ec};
  \node (direction_new)    [process, right of=threshold, xshift=3cm]    {Compute a\\ new direction};

  \draw[->]             (start) -- (initial);     
  \draw[->]             (initial) -- (next_0);   
  \draw[->]             (next_0) -- (decision_0);   
  \draw[->]      (decision_0) -- node {No} (set); 
  \draw[->]      (decision_0) -- node {Yes} (stop);
  \draw[->]      (initial) -- node[pos=0.5, fill=none](h) {} (next_0);
  \draw[->]             (return) -- node {Yes} (h); 
  \draw[->]      (set) -- node[pos=0.5, fill=none](h) {} (decision_1);
  \draw[->]             (return2zero) |- (h); 
  \draw[->]             (direction_new) |- (h);   
  \draw[->]      (decision_1) -- node {No} (position_new);
  \draw[->]      (decision_1) -- node {Yes} (transmitted);    
  \draw[->]      (position_new) -- (decision_2);         
  \draw[->]      (decision_2) -- node {No} (type); 
  \draw[->]      (decision_2) -- node {Yes} (backscattered);    
  \draw[->]      (type) -- (elastic);
  \draw[->]      (elastic) -- node {No} (angleANDenergy);
  \draw[->]      (elastic) -- node {Yes} (angle);   
  \draw[->]      (angle) |- ([shift={(0mm,-44mm)}]angle.south) -| (direction_new.south);
  \draw[->]      (angleANDenergy) -- (SE); 
  \draw[->]      (SE) -- node[pos=0.8, fill=none](h) {} (threshold);
  \draw[->]             (SEEyield) -| ([shift={(30mm,0mm)}]SEEyield.south) |- (h); 
  \draw[->]      (SE) -- node {Yes} (SEEyield);
  \draw[->]      (SE) -- node {No} (threshold);
  \draw[->]      (threshold) -- node {No} (direction_new);
  \draw[->]      (threshold) -- node {Yes} (absorbed);  
  \draw[->]      (absorbed) -| ([shift={(-50mm,0mm)}]absorbed.west) |- (return);
  \draw[->]      (transmitted) -| ([shift={(-9.35mm,0mm)}]transmitted.west) |- (return);
  \draw[->]      (backscattered) -| ([shift={(-5mm,0mm)}]backscattered.west) |- (return);
  \draw[->]      (return) -- node {No} (return2zero);
                                                                                                                                                                                                                                                                                                                                                                                                                                                                                                              
\end{tikzpicture}
\caption{Flow chart of the Monte Carlo program} \label{app:figure7}
\end{figure}

\newpage
\section{Definition of Coordinate System} \label{app:coord}
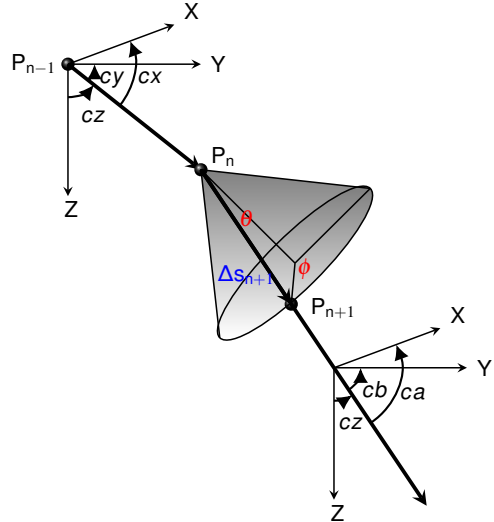
\begin{figure}[h]
\centering
\begin{tikzpicture}[thick,scale=0.35, every node/.style={font=\sffamily\sansmath\Huge, transform shape}]
\coordinate (O) at (0,0);
\coordinate (P) at (-5,4);

\shade[ball color = black] (-5,4) circle (0.25cm);
\draw[line width=0.2mm,->,>=stealth] (-5,4) -- (0,4) coordinate (B);
\draw[line width=0.2mm,->,>=stealth] (-5,4) -- (-5,-1) coordinate (C);
\draw[line width=0.2mm,->,>=stealth] (-5,4) -- (-1,5.5) coordinate (A);
\draw[line width=0.5mm,->,>=stealth] (-5,4) -- (0,0);

\draw[line width=0.2mm,->,>=stealth] (5,-7.5) coordinate (Q) -- (10,-7.5) coordinate (E);
\draw[line width=0.2mm,->,>=stealth] (5,-7.5) -- (5,-12.5) coordinate (F);
\draw[line width=0.2mm,->,>=stealth] (5,-7.5) -- (9,-6) coordinate (D);

\begin{scope}[rotate=45]
\shade[color = gray] (-4,-5) -- (0,0) -- (4,-5);
\shade[color = gray] (0,-5) circle(4cm and 1cm);
\shade[ball color = black] (0,0) circle (0.25cm);
\shade[ball color = black] (-1.2,-6) circle (0.25cm);
\draw[line width=0.2mm] (0,-5) -- (-1.2,-6);
\draw[line width=0.2mm] (0,-5) -- (4,-5);
\draw[line width=0.5mm,->,>=stealth] (0,0) -- (-1.2,-6);
\draw[line width=0.5mm,->,>=stealth] (0,0) -- (-3,-15) coordinate (R);
	\draw[line width=0.2mm] (0,-5) circle(4cm and 1cm);
	\draw[line width=0.2mm] (0,-5) -- (0,0);
	\draw[line width=0.2mm] (-4,-5) -- (0,0) -- (4,-5);
\end{scope}
\node[text width=6cm, anchor=east, left] at (5.5,6) {X};
\node[text width=6cm, anchor=east, left] at (6.5,4) {Y};
\node[text width=6cm, anchor=east, left] at (1,-1.5) {Z};
\node[text width=6cm, anchor=east, left] at (-1,4) {$\rm P_{n-1}$};
\node[text width=6cm, anchor=west, right] at (0.25,0.5) {$\rm P_{n}$};
\node[text=blue, text width=6cm, anchor=west, right] at (0.5,-4) {$\rm \Delta s_{n+1}$};
\node[text=red, text width=6cm, anchor=west, right] at (1.3,-1.85) {$\theta$};
\node[text=red, text width=6cm, anchor=west, right] at (3.5,-3.75) {$\phi$};
\node[text width=6cm, anchor=west, right] at (4,-5.25) {$\rm P_{n+1}$};
\node[text width=6cm, anchor=west, right] at (9.25,-5.5) {X};
\node[text width=6cm, anchor=west, right] at (10.25,-7.5) {Y};
\node[text width=6cm, anchor=west, right] at (4.75,-13) {Z};

\path pic[draw, ->, "$cx$", angle radius=2.5cm, angle eccentricity=1.25] {angle = O--P--A};
\path pic[draw, ->, "$cy$", angle radius=1cm, angle eccentricity=1.75] {angle = O--P--B};
\path pic[draw, ->, "$cz$", angle radius=1.25cm, angle eccentricity=1.75] {angle = C--P--O};

\path pic[draw, ->, "$ca$", angle radius=2.5cm, angle eccentricity=1.25] {angle = R--Q--D};
\path pic[draw, ->, "$cb$", angle radius=1cm, angle eccentricity=1.75] {angle = R--Q--E};
\path pic[draw, ->, "$cz$", angle radius=1.25cm, angle eccentricity=1.75] {angle = F--Q--R};
\end{tikzpicture}
\caption{Definition of coordinate system used in the Monte Carlo simulation program} \label{app:figure6}
\end{figure}

The basic geometry for the simulation assumes that the electron undergoes an elastic scattering event at some point $P\textsubscript{n}$, having traveled to $P\textsubscript{n}$ from a previous scattering event at $P\textsubscript{n-1}$ as shown in Figure \ref{app:figure6}.
To calculate the position of the new scattering point $P\textsubscript{n+1}$, we first require to know the distance $\Delta s_{n+1}$ between $P\textsubscript{n+1}$ and the preceding point $P\textsubscript{n}$.

The path is described using direction cosines, ca, cb and cc. The coordinates at the end of the step at $P\textsubscript{n+1}$, $x\textsubscript{n+1}$, $y\textsubscript{n+1}$ and $z\textsubscript{n+1}$, are then related to the coordinates $x\textsubscript{n}$, $y\textsubscript{n}$ and $z\textsubscript{n}$ at $P\textsubscript{n}$ by the formulas \cite{heinrich1976use}
\begin{equation*}
\begin{aligned}
& x_{n+1} = x_{n}+\Delta s_{n+1} \cdot ca \\
& y_{n+1} = y_{n}+\Delta s_{n+1} \cdot cb \\
& z_{n+1} = z_{n}+\Delta s_{n+1} \cdot cc
\end{aligned}
\end{equation*}
The direction cosines $ca$, $cb$, $cc$ are found from the direction cosines $cx$, $cy$ and $cz$ with which the electron reached P\textsubscript{n}. The result is
\begin{equation*}
\begin{aligned}
& ca = (cx \cdot \cos \theta)+(V1 \cdot V3)+(cy \cdot V2 \cdot V4) \\
& cb = (cy \cdot \cos \theta)+(V4 \cdot (cz \cdot V1 - cx \cdot V2 )) \\
& cc = (cz \cdot \cos \theta)+(V2 \cdot V3)-(cy \cdot V1 \cdot V4)
\end{aligned}
\end{equation*}
where
\begin{equation*}
\begin{aligned}
& V1 = AN \cdot \sin \theta \\
& V2 = AM \cdot AN \sin \theta \\
& V3 = \cos \phi \\
& V4 = \sin \phi
\end{aligned}
\end{equation*}
and 
\begin{equation*}
\begin{aligned}
& AM = -\frac{cx}{cz} \\
& AN = \frac{1}{\sqrt{1+AM \cdot AM}}
\end{aligned}
\end{equation*}

\end{document}